\documentclass[11pt]{article}
\usepackage[margin=1in,letterpaper]{geometry}
\usepackage{amsmath}
\usepackage{amssymb}
\usepackage{setspace}
\usepackage{cite}
\usepackage{hyperref}
\usepackage{color}
\usepackage[pdftex]{graphicx}
\graphicspath{{./figure/}}
\usepackage[vlined,linesnumbered,boxed]{algorithm2e}
\usepackage{ctable}
\usepackage{multirow}
\usepackage[symbol*]{footmisc}
\usepackage{fourier-orns}
\def\Vhrulefill{\leavevmode\leaders\hrule height 0.6ex depth\dimexpr0.3pt-0.6ex\hfill\kern0pt}

\usepackage{fixltx2e}

\usepackage{pdflscape}

\usepackage[affil-it]{authblk}

\usepackage{fancyhdr}
\fancypagestyle{mystyle}{%
\fancyhf{}
\fancyhead[LE,RO]{\small \thepage}
\fancyhead[LO,RE]{\small \scshape \nouppercase{\leftmark}}
}
\newcommand{\citep}[1]{\cite{#1}}
\newcommand{\citealp}[1]{\cite{#1}}

\newcommand{\mylinespacing}{1.29}
\newcommand{\mytablespacing}{1}

\newcommand{\eg}{{\it e.g.}}
\newcommand{\ie}{{\it i.e.}}

\newcommand{\gyes}{{\textsc{G}$_{+}$}}
\newcommand{\gno}{{\textsc{G}$_{0}$}}

\newcommand{\butil}{\textsc{Butil}}
\newcommand{\bqlen}{\textsc{WBlen}}
\newcommand{\pfault}{\textsc{Pfault}}
\newcommand{\wsize}{\textsc{Wsize}}
\newcommand{\rsize}{\textsc{Rsize}}
\newcommand{\wiops}{\textsc{Wiops}}
\newcommand{\riops}{\textsc{Riops}}
\newcommand{\car}{\textsc{CAR}}

\newcommand{\pone}{\textbf{P1}}
\newcommand{\ptwo}{\textbf{P2}}
\newcommand{\pthree}{\textbf{P3}}
\newcommand{\pfour}{\textbf{P4}}
\newcommand{\pfive}{\textbf{P5}}

\newcommand{\ba}{\textbf{a}}
\newcommand{\bb}{\textbf{b}}
\newcommand{\bc}{\textbf{c}}
\newcommand{\bd}{\textbf{d}}
\newcommand{\be}{\textbf{e}}

\begin{document}
\pagestyle{plain}

\noindent
{\Large\textbf{Will solid-state drives accelerate your bioinformatics?\\In-depth profiling, performance analysis, and beyond}}
\vspace{1em}

\noindent
Sungmin Lee$^1$, Hyeyoung Min$^2$, and Sungroh Yoon$^1$\footnote{To whom correspondence should be addressed: \href{mailto:sryoon@snu.ac.kr}{sryoon@snu.ac.kr}}


\vspace{1em}

\noindent
\emph{\small{$^1$Electrical and Computer Engineering, Seoul National University, Seoul 151-744, Korea}}\\
\emph{\small{$^2$College of Pharmacy, Chung-Ang University, Seoul 156-756, Korea}}

\vspace{1ex}

\begin{spacing}{\mylinespacing}

\begin{center}
\begin{minipage}{0.895\linewidth}
\small
\textbf{Abstract}\hspace{0.2cm}\hrulefill

\vspace{1ex}
\noindent
A wide variety of large-scale data has been produced in bioinformatics. In response, the need for efficient handling of biomedical big data has been partly met by parallel computing. However, the time demand of many bioinformatics programs still remains high for large-scale practical uses due to factors that hinder acceleration by parallelization. Recently, new generations of storage devices have emerged, such as NAND flash-based solid-state drives (SSDs), and with the renewed interest in near-data processing, they are increasingly becoming acceleration methods that can accompany parallel processing. In certain cases, a simple drop-in replacement of hard disk drives (HDDs) by SSDs results in dramatic speedup. Despite the various advantages and continuous cost reduction of SSDs, there has been little review of SSD-based profiling and performance exploration of important but time-consuming bioinformatics programs. For an informative review, we perform in-depth profiling and analysis of 23 key bioinformatics programs using multiple types of devices. Based on the insight we obtain from this research, we further discuss issues related to design and optimize bioinformatics algorithms and pipelines to fully exploit SSDs. The programs we profile cover traditional and emerging areas of importance, such as alignment, assembly, mapping, expression analysis, variant calling, and metagenomics. We explain how acceleration by parallelization can be combined with SSDs for improved performance and also how using SSDs can expedite important bioinformatics pipelines, such as variant calling by the Genome Analysis Toolkit (GATK) and transcriptome analysis using RNA sequencing (RNA-seq). We hope that this review can provide useful directions and tips to accompany future bioinformatics algorithm design procedures that properly consider new generations of powerful storage devices.

\textbf{Availability:} \href{http://best.snu.ac.kr/pub/biossd}{http://best.snu.ac.kr/pub/biossd}

\Vhrulefill
\end{minipage}
\end{center}


\section{Introduction}\label{s:introduction}

Enabled by breakthroughs in data generation, collection, and analysis technologies, we are living in the era of big data\citep{schadt2012changing}. Novel data-driven research and business opportunities are envisioned in many disciplines, and biomedicine is not an exception. The recent trend toward personalized precision medicine has triggered the accumulation of a great deal of biomedical data from various sources~\citep{holzinger2014biomedical}, such as (epi-/meta-/pharmaco-)genomics, transcriptomics, proteomics, metabolomics, wearable mobile devices, and crowd-sourced scientific games\citep{lee2014rna}.


The need for efficient processing of biomedical big data has been partly met by parallel computing that spans from shared-memory machines (e.g., multicore CPUs and GPUs) to distributed systems (e.g., MPI/Hadoop/Spark-based cloud computing). For instance, the Broad Institute and Intel Corporation have been jointly working on parallelizing the Genome Analysis Toolkit (GATK, \citealp{mckenna2010genome}). Its sequential implementation takes more than 360 hours to genotype a single personal human genome, but this collaboration recently reported that it is possible to gain a more than 10-fold speedup by employing multicore processors.


Nevertheless, the time demand of many bioinformatics programs still remains unsatisfactory for large-scale practical uses, due to various reasons that hinder acceleration by parallelization, such as limited parallelism in the algorithm, frequent data transfers among computing units, and high cost (time and resources) of parallelization. Additional methods for acceleration (other than parallel computing) have been sought, including storage-centric approaches that are emerging with the renewed interest in near-data processing~\citep{balasubramonian2014near}.


Traditionally, there has been a substantial difference between the pace of improvements in CPUs and storage technologies, also known as the CPU-IO performance gap~\citep{katz1989disk}. With the advent of NAND flash-based solid-state drives (SSDs), this gap is becoming narrower than ever, along with the gradual transition to fast host interfaces (such as PCI Express). SSDs show substantially higher performance than hard disk drives (HDDs) especially when there are frequent random input-output (IO) requests~\citep{seo2014io}, not to mention their mechanical advantages originating from the lack of moving internal components. In data science and engineering, various workloads with abundant random IOs have been successfully accelerated often by a simple ‘drop-in’ replacement of HDDs by SSDs. Furthermore, traditional data analytics algorithms are being redesigned to fully exploit the new, fast secondary storage~\citep{lee2008case}.


Despite the simplicity (e.g., drop-in replacement without any other modifications) and continuous cost reduction fostering widespread use of SSDs, there has been little review of SSD-based profiling and performance exploration in the bioinformatics community. In this review, we compare the performance of 23 well-known bioinformatics programs (see Table~\ref{t:tool_list}) using multiple types of SSDs and HDDs. The programs we analyze cover traditional and emerging bioinformatics areas of high importance, such as sequence alignment, genome assembly, read mapping, gene expression analysis, motif finding, variant calling, and metagenomics. We classify these bioinformatics tools into two groups, depending on the effectiveness of SSDs on speedup, and investigate the factors that cause the difference from a storage system perspective.

%

Based on the insight obtained from the research, we further discuss issues in implementing and selecting bioinformatics algorithms and pipelines with the SSDs under consideration. For instance, we show that acceleration by parallelization can be accompanied by SSDs to yield extra runtime improvements. Examples include ABySS\citep{simpson2009abyss} (a parallel short-read assembler) and the GATK (which uses the MapReduce framework~\citep{dean2008mapreduce}). In our experiments, ABySS and a variant-calling pipeline using the GATK achieved 51.7 and 35.7 times speedup, respectively, when using SSDs. Another discussion on SSD-based acceleration comes from the short-read aligners for next-generation sequencing (NGS)~\citep{metzker2009sequencing}. We compare Maq~\citep{li2008mapping}, Burrows-Wheeler Aligner (BWA, \citealp{li2009fast}), and Bowtie 2~\citep{langmead2012fast} in terms of runtime and quality metrics before and after using SSDs and analyze the result from storage-system perspectives. Based on this analysis, we further discuss how to assess alternative bioinformatics programs in terms of the viability of SSD-based acceleration.

To the best of the authors’ knowledge, this review presents the first in-depth profiling analysis of major bioinformatics programs targeted at revealing opportunities and limitations of using SSDs for acceleration of bioinformatics tools. We hope that this review can provide useful directions and tips that should accompany future bioinformatics algorithm design procedures that properly consider new generations of powerful storage devices. 

\begin{landscape}
\begin{spacing}{\mytablespacing}
	
\ctable[
	caption = {List of the twenty three bioinformatics programs profiled and analyzed in this work},
	label = {t:tool_list},
	doinside = \small
]{llllll}{
\tnote[]{\gyes, programs with 2x or more speedup; \gno, programs with negligible improvements; MSA, multiple sequence alignment; NJ, neighbor joining; HMM, hidden Markov model; EM, expectation maximization; $\dagger$ speedup by Intel 520 SSD over Seagate Barracuda HDD [see Tables~\ref{t:ssd_spec} and \ref{t:hdd_spec} for specifications and Table~\ref{t:data_list} for input data].}
}{
\toprule 
& Name  &Task &Main algorithm& Source & Speedup$^\dagger$\\ 
\midrule 
\gyes & GATK BaseRecal & Base quality recalibration& generates recalibration table based on covariates & \citealp{mckenna2010genome} & 78.4 \\
 & Samtools  &Utility tool& sorting, merging, indexing large sequence alignment& \citealp{li2009sequence} & 77.2 \\
 & ABySS  & NGS assembler& distributed de Bruijn graph, hash table searching &\citealp{simpson2009abyss} & 51.7\\
 & Cluster3 &  Microarray analsysis& calculating pairwise sequence distance, clustering& \citealp{de2004open} & 50.0\\
 & Blat  &  Sequence alignment & index searching on non-overlapping k-mers& \citealp{kent2002blat}& 23.6 \\
 & Reptile & NGS denoising& MSA with Hamming distance, k-spectrum extraction& \citealp{yang2010reptile}  & 13.7\\
 & GATK Aligner & Sequence realignment&Smith-Waterman local realignment&\citealp{mckenna2010genome} & 12.6\\
 & Maq & NGS assembler& ungapped sequence alignment, maximizing posterior probability& \citealp{li2008mapping} & 10.1\\
 & Tophat & RNA-seq analysis&segmented sequence alignment using Bowtie& \citealp{trapnell2009tophat} & 3.2\\
 & MC-UPGMA& Microarray analsysis& memory-constrained multi-round hierarchical clustering& \citealp{loewenstein2008efficient}  &2.7\\
\midrule 
\gno & BWA &  Sequence alignment&Burrows-Wheeler transform, trie traversal& \citealp{li2009fast} &1.09\\
 & Blast & Sequence alignment&seed-based local sequence alignment& \citealp{altschul1990basic} & 1.08\\
 & ClustalW & Sequence alignment&multiple sequence alignment using NJ guide tree& \citealp{thompson1994clustal}& 1.06\\
 & GATK Unified &Genome variant calling& Bayesian likelihood modeling& \citealp{mckenna2010genome} & 1.05\\
 & GATK PrintReads &Utility tool&sorting, and merging sequence alignments& \citealp{mckenna2010genome} & 1.03\\
 & Scripture &RNA-seq analysis&sequence alignment using TopHat, graph traversal& \citealp{guttman2010ab} & 1.02\\
 & IGVtools &Utility tool&sequence alignment indexing, sorting& \citealp{robinson2011integrative} & 1.02\\
 & Meme &Motif finding&expectation-maximization, greedy search& \citealp{bailey2006meme} & 1.00\\
 & Bowtie 2 &Sequence alignment&Burrows-Wheeler-based sequence alignment& \citealp{langmead2012fast} & 1.00\\ 
 & Mosdi &Motif finding&HMM-based statistical modeling, suffix tree traversal& \citealp{marschall2009efficient} & 1.00\\
 & AmpliconNoise &NGS denoising&Needleman-Wunsch, hierarchical clustering, EM& \citealp{quince2011removing} &1.00\\
 & Weeder &Motif finding&suffix tree-based exhaustive searching& \citealp{pavesi2007weederh} & 1.00\\
 & ErmineJ &Microarray analsysis&permutation, rank-based statistics analysis& \citealp{gillis2010gene} &0.97\\
\bottomrule
}
\end{spacing}

\end{landscape}

\pagestyle{mystyle}

\section{Results: SSD-leveraged Acceleration}\label{s:result}

\subsection{SSD-leveraged resurrection of hash-based aligners}
As a warm-up case, we tested how using SSDs can accelerate well-known bioinformatics programs simply by the drop-in replacement of HDDs by SSDs in the same computer without any other modifications in hardware or software. To this end, we used the short-read alignment tools for next-generation sequencing~\citep{metzker2009sequencing}. Note that the first wave of such tools, mostly hash-based methods (\eg, Maq), has been gradually replaced by Burrows-Wheeler Transform (BWT) based methods (\eg, Bowtie 2 and BWA), mainly because of their rapid searching capabilities backed by smaller memory footprints, albeit a sacrifice in accuracy~\citep{flicek2009sense}.

\begin{figure}
\includegraphics[width=\linewidth]{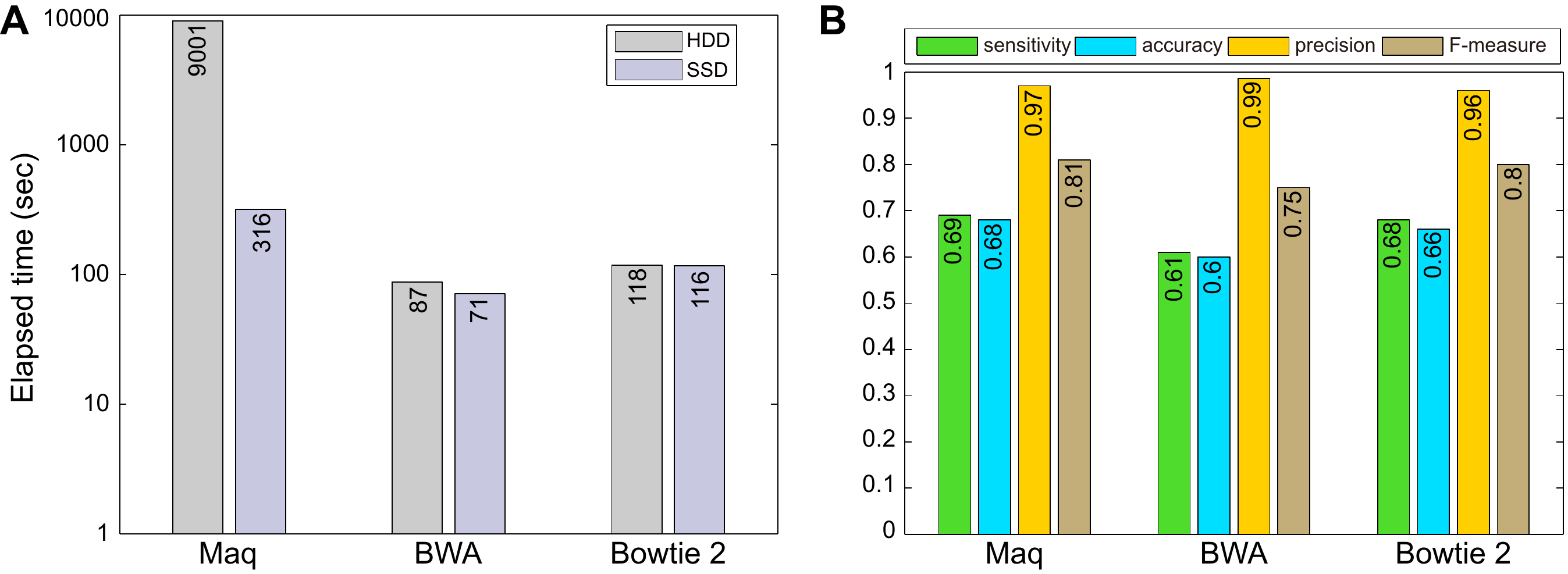}
\caption{Performance comparison of three short-read aligners: Maq~\citep{li2008mapping}, BWA~\citep{li2009fast}, and Bowtie~2~\citep{langmead2012fast}. (\ba) Runtime. (\bb) Quality measured in sensitivity, accuracy, precision, and F-measure. [SSD: Samsung 840 Pro (128GB), HDD: Seagate Barracuda (1TB, 7200rpm), data: \emph{Staphyloccus aureus} whole genome sequence~\citep{salzberg2012gage}]}\label{f:ssd_revives}
\end{figure}

\figurename~\ref{f:ssd_revives} shows the running time and quality of Maq, BWA, and Bowtie 2. Refer to the figure caption for details of the devices and the data set used. As expected, when HDDs are used, the runtime of Maq is significantly higher than that of Bowtie 2 or BWA. Maq is a hash-based method while Bowtie 2 and BWA are more memory-efficient BWT-based techniques. Consequently, these second-generation methods usually run faster than the first-generation aligners especially when the data size is large and swaps frequently occur. When SSDs are used, Maq is still the slowest, but the runtime gap has become dramatically narrower, leveraged by the enhanced IO performance and reduced swap cost of SSDs.

Given this boost in runtime and the advantage in quality measured using various metrics as shown in \figurename~\ref{f:ssd_revives}(b), it would be possible to use Maq instead of Bowtie 2 or BWA when high values for quality metrics are desired. A simple drop-in replacement of HDDs by SSDs has made the earlier generation of tools competitive to the later generation of tools to some extent.

\subsection{Measuring speedup of bioinformatics programs}\label{ss:first_look}
To further investigate what kind of bioinformatics tools can be accelerated by using SSDs, we prepared a total of 23 bioinformatics programs listed in Table~\ref{t:tool_list} and measured the speedup by drop-in replacements of HDDs by SSDs. Refer to Tables~\ref{t:ssd_spec} and \ref{t:hdd_spec} in Section~\ref{ss:exp_setup} for more details of the experiments.

\begin{figure}
\centering
\includegraphics[width=0.8\linewidth]{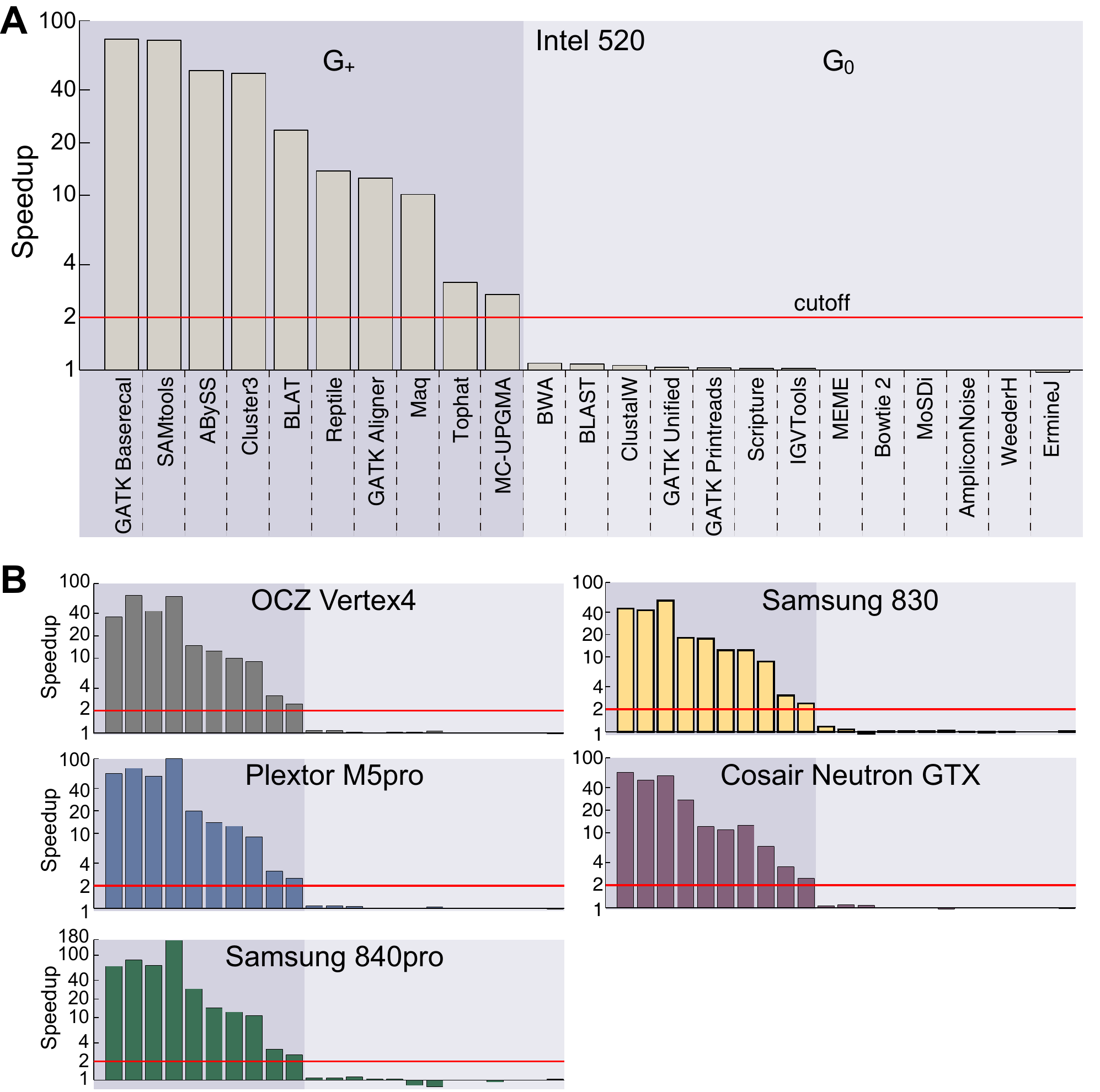}
\caption{Speedup of 23 bioinformatics programs by drop-in replacements of HDDs by SSDs. [\gyes, programs with 2x or more speedup; \gno, programs with negligible improvements] (\ba) SSD: Intel 520 (120GB), HDD: Seagate Barracuda (1TB, 7200rpm). (\bb) SSD: five different models listed in Table~\ref{t:ssd_spec}, HDD: the same as in (a). The order of the programs placed below the x-axis remains the same as in (a). Additional results form comparing a complete set of SSD-HDD pairs is available at http://best.snu.ac.kr/pub/biossd.}\label{f:speedup}
\end{figure}



%
%
%
%


The result is shown in \figurename~\ref{f:speedup}. Using SSDs yielded substantial speedup for certain programs (\eg, GATK BaseRecal) but was not always effective. Regardless of the specific SSD used for measurement, we were able to divide the 23 programs into two groups, namely \gyes~(the programs with 2x or more speedup) and \gno~(the programs with negligible or no improvements). The programs in each of these two groups are listed in \tablename~\ref{t:tool_list}. To find the root-cause reason that separates these two groups, we will further profile and analyze these 23 programs from storage system perspectives in Section~\ref{s:analysis}.

%


Note that the result shown in \figurename~\ref{f:speedup}(a) is from using a 120GB Intel 520 SSD in place of a 1TB Seagate Barracuda HDD (3.5 inch). The results from the other five SSDs are shown in \figurename~\ref{f:speedup}(b). Using different SSDs and HDDs did not change the group membership of each program but only its speedup ranking within each group. In what follows, we thus present the results obtained from using an Intel 520 and a Seagate Barracuda unless otherwise stated. The results from using the other combinations of SSDs and HDDs are available online at http://best.snu.ac.kr/pub/biossd.

\subsection{Accelerating bioinformatics pipelines by SSDs}
\begin{landscape}
	
\begin{figure*}
\centering
\includegraphics[width=\linewidth]{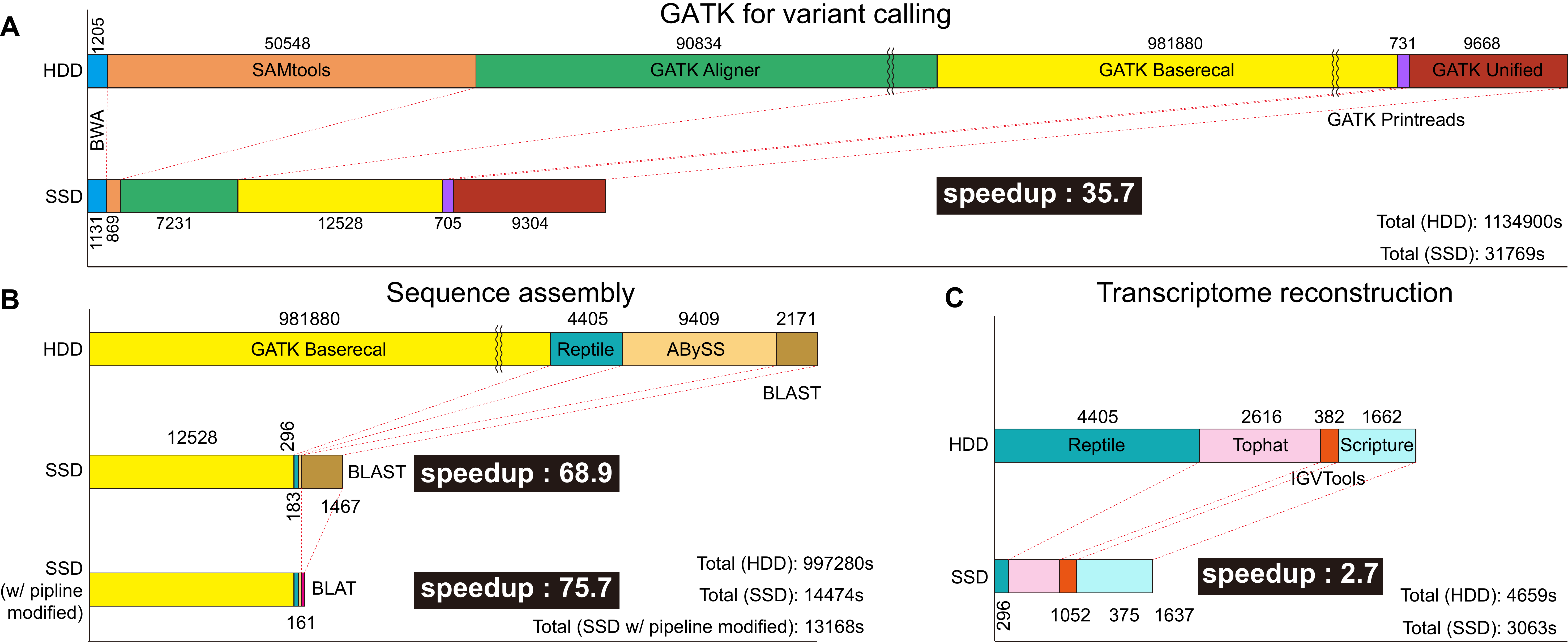}
\caption{SSD-based acceleration of bioinformatics pipelines. (\ba) variant calling by GATK~\citep{mckenna2010genome}. [data: NA12878 human whole genome sequence~\citep{10002010map}] (\bb) Sequence assembly and annotation~\citep{birol2009novo}. [data: \emph{Staphyloccus aureus} whole genome sequence~\citep{salzberg2012gage}] (\bc) Transcriptome reconstruction~\citep{guttman2010ab}. [data: Mouse (mm9) reads (see Table~\ref{t:data_list} for a link)]}\label{f:pipeline}
\end{figure*}

\end{landscape}

Based on the initial profiling results described in Section~\ref{ss:first_look}, we further tested if there is any performance gain by using SSDs for running a bioinformatics pipeline that consists of multiple component programs. As shown in \figurename~\ref{f:pipeline}, we measured the runtime of three bioinformatics pipelines before and after a drop-in replacement of HDDs by SSDs. The pipelines analyzed are for variant calling by the Genome Analysis Toolkit (GATK)~\citep{mckenna2010genome}, whole-genome sequence assembly and annotation~\citep{birol2009novo}, and transcriptome reconstruction~\citep{guttman2010ab}.

\figurename~\ref{f:pipeline}(a) illustrates the breakdown of the runtime of the GATK pipeline for SNP calling. The pipeline consists of the component tools for sequence alignment and formatting using BWA~\citep{li2009fast} and Samtools~\citep{li2009sequence}, sequence realignment (GATK Aligner), sequence base-quality recalibration (GATK BaseRecal), result merge (GATK PrintReads), and SNP and indel calling (GATK Unified). By a simple drop-in replacement, we could achieve more than a 35 times decrease in the runtime of the whole pipeline. The majority of the speedup is due to the reduced runtime of formatting (Samtools, 77.2x speedup), sequence realignment (GATK Aligner, 12.6x speedup), and base-quality recalibration (GATK BaseRecal, 78.4x speedup). %

The second pipeline depicted in \figurename~\ref{f:pipeline}(b) carries out sequence assembly and annotation. The first three steps account for most of the improvements and consist of GATK Baserecal, Reptile~\citep{yang2010reptile}, and ABySS~\citep{simpson2009abyss}, which are all accelerated significantly by SSDs, according to Table~\ref{t:tool_list}. Replacing Blast with Blat gave additional runtime reduction, producing 75.7x total speedup over HDDs. Of note is that ABySS, a parallel short-read assembler, got boosted more than 50 times by SSDs. This is an example in which combination of computing parallelization and SSD-based storage can yield a dramatic performance gain.

\figurename~\ref{f:pipeline}(c) shows the third pipeline for transcriptome reconstruction \citep{guttman2010ab} in RNA-seq experiments~\citep{mortazavi2008mapping}. The amount of speedup was smaller than the above two. Although the most time-consuming step (Reptile) of the pipeline was accelerated significantly by SSDs, the total runtime of the pipeline was relatively shorter, and the effect of runtime reduction in Reptile got eclipsed by the Scripture~\citep{guttman2010ab} step. We expect that using a larger data set will reveal the effect of SSD-based runtime reduction. (Related results are presented in Section~\ref{ss:additional}.) 

\section{Results: Profiling and Analysis}\label{s:analysis}
This section elaborates how we profiled and analyzed the 23 bioinformatics programs under study. We first measured important storage features for each program and then clustered the programs with respect to the measured feature values. The measurement and clustering allowed us to discover IO patterns that can not only differentiate \gyes~and \gno~but also provide useful insight into when SSDs can be effective for acceleration and when not.

\subsection{Measuring storage features}\label{ss:measuring_storage_features}
For each of the 23 bioinformatics programs, we measured eight features that are widely used in storage research. Table~\ref{t:io_features} lists more details of these features and their acronyms to be used in the paper. Using these features, we will consider the randomness and the amount of IOs involved in these 23 programs. The amount of IOs is measured by \butil, \riops, \wiops, and \pfault, whereas the IO randomness is measured by \car, \rsize, \wsize, and \bqlen. More details can be found in Section~\ref{ss:feature_exp}.


\begin{spacing}{\mytablespacing}
\ctable[
	caption = {List of storage features used (see Section~\ref{ss:feature_exp} for details)},
	label = {t:io_features},
	pos = t,
]{lll}{
}{
\toprule
ID & Feature & When high (low)\\
\midrule
\butil & interface bandwidth utilization & large (small) data transfers\\
\riops & read IO per second & (in)frequent reads \\
\wiops & write IO per second & (in)frequent writes\\
\pfault & \# page faults per second & (in)frequent page swaps\\
\car & Consecutive Access Ratio & sequential (random) access \\
\rsize & read size per request & sequential (random) reads\\
\wsize & write size per request & sequential (random) writes\\
\bqlen & write buffer length & many (few) writes in queue\\
\bottomrule
}
\end{spacing}


\begin{figure}
\centering
\includegraphics[width=0.46\linewidth]{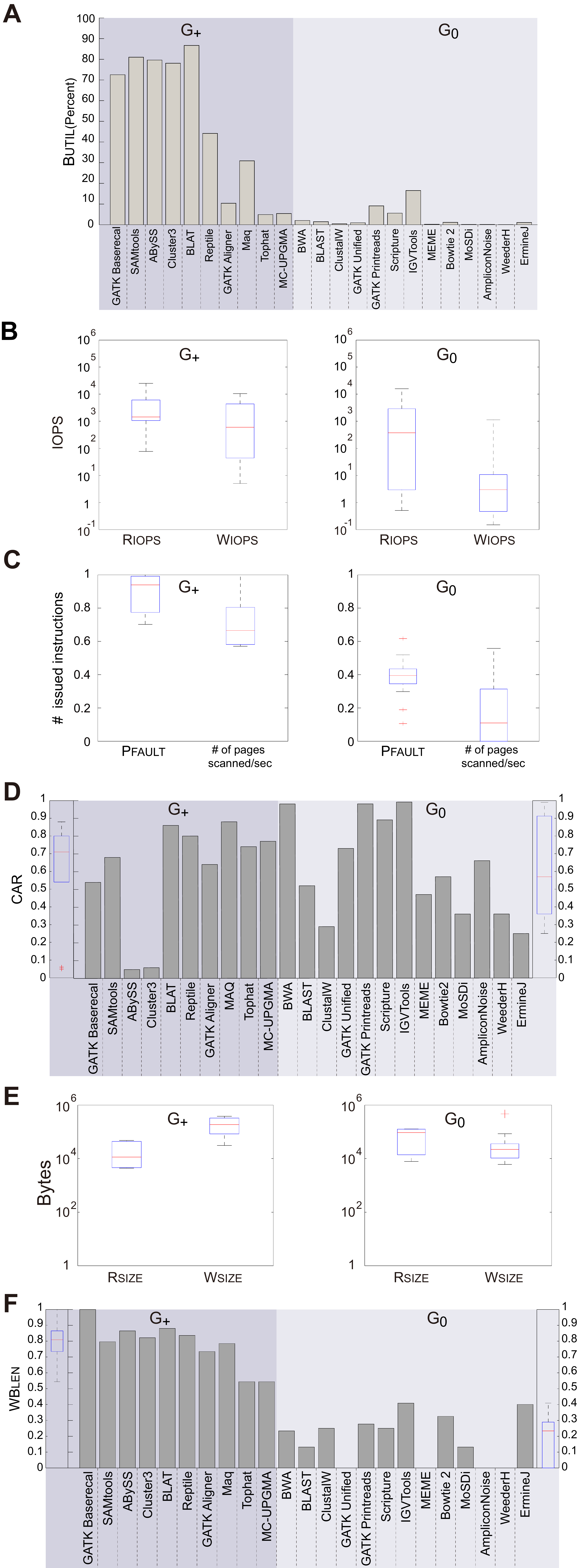}
\caption{Storage feature measurements. (\ba) Bandwidth utilization of host-storage interface (\butil). (\bb) Read IOPS (\riops) and write IOPS (\wiops). (\bc) The number of page faults per second (\pfault). (\bd) Consecutive Access Ratio (\car). (\be) Read size per request (\rsize) and write size per request (\wsize). (\textbf{f}) Storage buffer queue length (\bqlen). [see Table~\ref{t:io_features} and Section~\ref{ss:feature_exp} for more details of these features]}\label{f:charac}
\end{figure}

The measurement results are shown in \figurename~\ref{f:charac}. Overall, we can make the following observations:
\begin{itemize}
\item[\textbf{O1}] The features related to the number of IO operations issued by the host (\riops~and \wiops) have higher values for \gyes.
\item[\textbf{O2}] The features related to the amount or frequency of transfers between the host memory and the storage (\butil~and \pfault) are higher for \gyes. 
\item[\textbf{O3}] Each of the features related to IO randomness (\rsize, \wsize, and \car) shows a different pattern: \rsize~is higher for \gno~(\ie, negligible speedup for programs with many sequential reads), \wsize~is higher for \gyes~(\ie, notable speedup for programs with many random writes), and \car~is moderately higher for \gyes.
\item[\textbf{O4}] The feature affected by both the amount of data transfers and IO randomness (\bqlen) is consistently higher for \gyes.
\end{itemize}

\textbf{O1} and \textbf{O2} can be explained by the fact that SSDs normally support higher IOPS while incurring less overheads for swaps. Thus, the programs with more IO operations and page faults can be more effectively accelerated by SSDs. 
\textbf{O3} and \textbf{O4} are related to the fact that SSDs are superior for handling random IOs, but part of these observations is not completely intuitive at first. 

For instance, not only SSDs but also HDDs normally have DRAM buffers that can hide latency incurred by random writes, implying that programs with many random writes will not see significant speedup by using SSDs. This implication is seemingly against \textbf{O3}. In addition, according to \textbf{O3}, \car~is higher for \gyes, which seems to suggest that the programs in \gyes~show less randomness. Given that SSDs are effective for handling random IOs, \textbf{O3} is seemingly inconsistent with the fact that the programs in \gyes~are accelerated more by using SSDs. Section~\ref{ss:impact_randomness_on_speedup} include further explanations of \textbf{O3} and \textbf{O4} that can answer these riddles.

\subsection{Pattern discovery by clustering}
\begin{figure}
\centering
\includegraphics[width=0.7\linewidth]{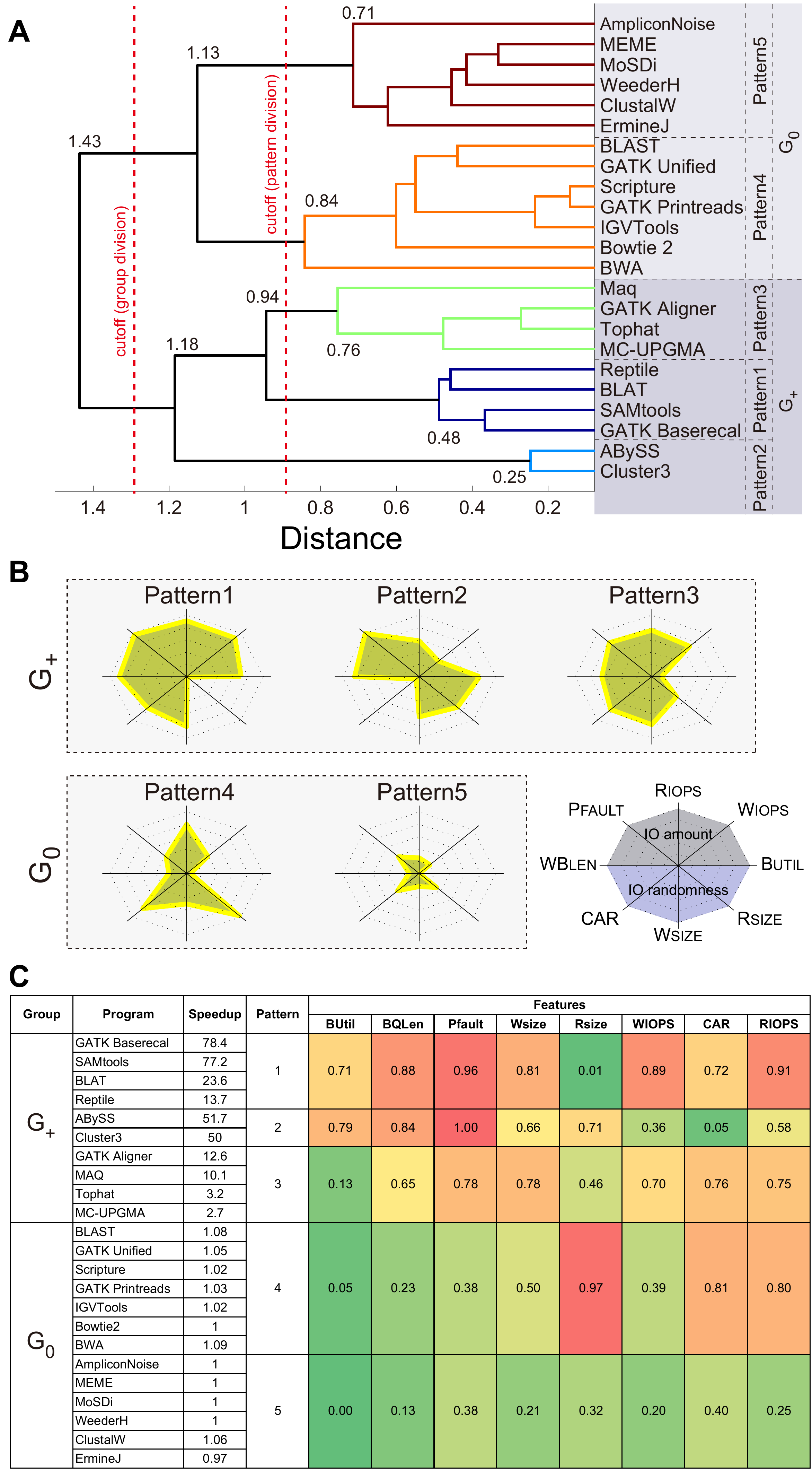}
\caption{Clustering bioinformatics programs based on the eight storage features listed in Table~\ref{t:io_features}. (\ba) Dendrogram and pattern definitions. The numbers represent the pairwise distance. (\bb) Radar chart representations of the average feature values for each pattern. Legend is also shown. (\bc) The numberical values of the average features depicted on the axes of the radar charts in (b). The names and the speedup amounts are also presented.}\label{f:clustering}
\end{figure}

Observations \textbf{O1}--\textbf{O4} only reveal overall trends. For a specific program, the prediction of the effectiveness of SSD simply using individual storage features alone may not be accurate. For example, some programs in \gno~have high \butil, \riops, and \wiops~but do not show significant speedup. To see the combinations of features leading to effective speedup and to find patterns that can help grouping bioinformatics programs in terms of IO behavior, we tried clustering the 23 programs based on the eight storage features.

\figurename~\ref{f:clustering}(a) shows the dendrogram obtained by agglomerative hierarchical clustering with the average linkage. We use the Euclidean distance metric to measure the distance between two vectors, each of which consists of the eight measurement values normalized and ranged in $[0,1]$ (see Section~\ref{ss:exp_setup}). Cutting the dendrogram near the root bifurcation point reveals the two groups \gyes~and \gno. Cutting it at the smaller distance as shown in the plot produces five clusters or \emph{patterns}. Group \gyes~consists of three patterns (denoted by \textbf{P1}, \textbf{P2}, and \textbf{P3}), while group \gno~contains two (denoted by \textbf{P4} and \textbf{P5}).

\begin{figure*}
\centering
\includegraphics[width=\linewidth]{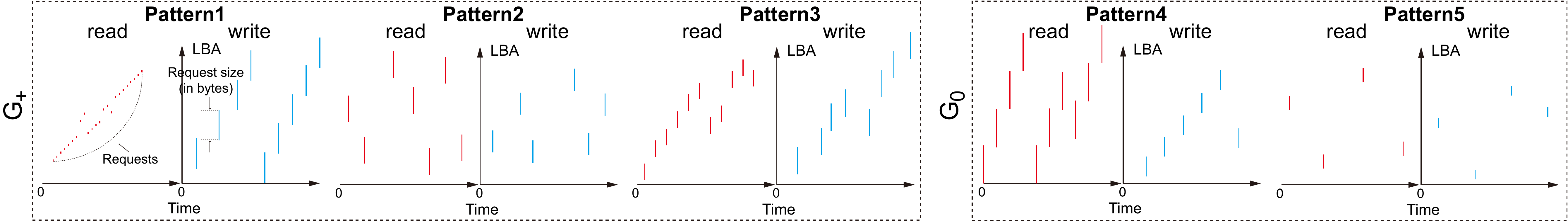}
\caption{Representative IO traces for each of the five patterns shown in \figurename~\ref{f:clustering}. A vertical bar corresponds to an IO request, and its length represents the read or write size. The quantity of vars in each plot is proportional to the IO amount, while the distribution of accessed LBAs represents the IO randomness. [LBA, logical block address; \gyes, programs with 2x or more speedup; \gno, programs with negligible or no improvements]}\label{f:lba_map}
\end{figure*}

\figurename~\ref{f:clustering}(b) shows the radar chart representation of the average feature values for each pattern. \figurename~\ref{f:clustering}(c) shows the numerical values depicted in the radar charts. Evidently, the most notable difference between the three patterns in \gyes~and the two patterns in \gno~is the average \pfault~value. However, the effect of \pfault~may not be observed clearly when the main memory is large, and we need to compare different patterns using other storage features.

To facilitate the comparison of the five patterns discovered, we present their representative IO traces in \figurename~\ref{f:lba_map}. We show two traces (read and write) for each pattern. In each trace, the x-axis and the y-axis represent the IO request time and the logical block address (LBA), respectively. Each vertical line corresponds to an IO request, and its length matches the read/write size. 

Using the information presented in \figurename~\ref{f:clustering} and \ref{f:lba_map}, we can identify notable characteristics of each pattern. For instance, \pone~has a high amount of IOs, frequent random reads and sequential writes. \pone~shows the lowest \rsize~(0.01) among all the five patterns, meaning that the read size per request is very small. Additionally, a \car~of 0.72 suggests that 72\% of the IO requests make consecutive access to the LBA. Taken together, we expect small data reads from often consecutive locations. In contrast, \wsize~(0.81) of \pone~is the highest among all the patterns. Again with 72\% \car, this implies frequent sequential writes of relatively large data. \riops~and \wiops~are the highest in \pone, implying a high amount of IOs. This is also backed by the high values of \butil, \bqlen, and \pfault. In particular, high \wiops~is responsible for high \bqlen. 

In a similar manner, we can also interpret the other patterns.

\subsection{Impact of IO randomness on speedup}\label{ss:impact_randomness_on_speedup}

We present how the IO randomness affects the amount of speedup by SSDs. We also show that the randomness alone may not always be a good indicator of speedup and should be accompanied by other storage features for more accurate prediction. 

\begin{figure*}
\centering
\includegraphics[width=\linewidth]{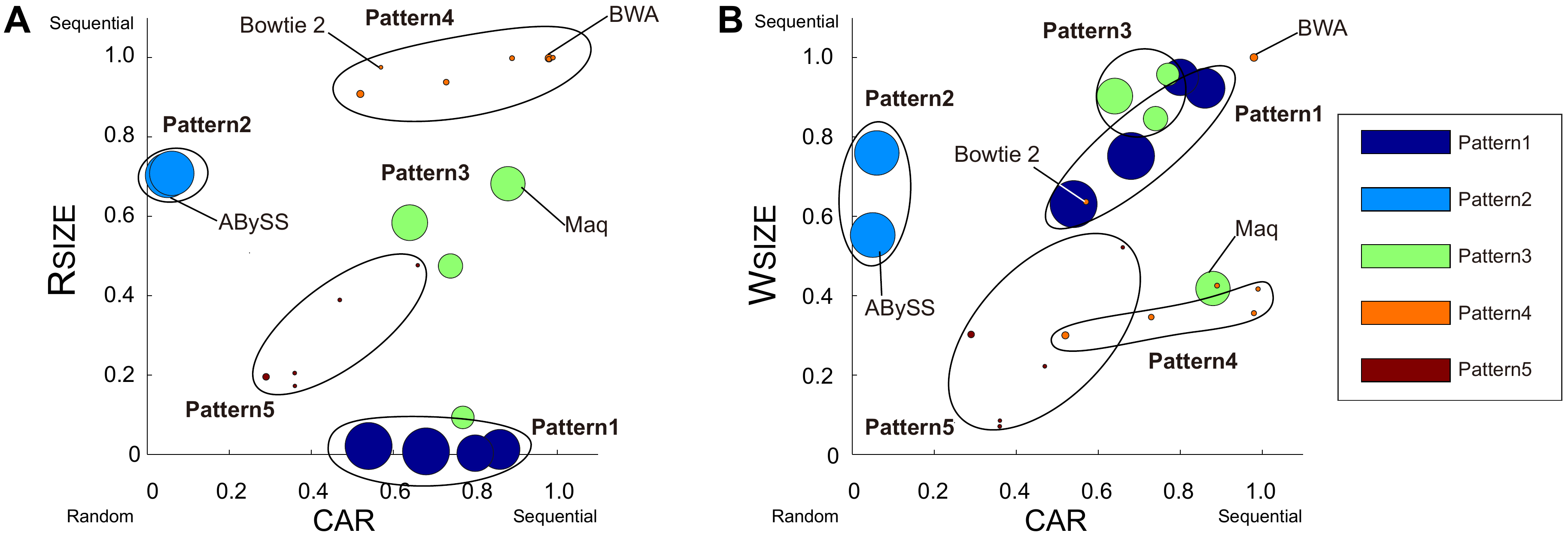}
\caption{Impact of randomness on speedup. Each circle represents one of the 23 bioinformatics programs listed in Table~\ref{t:tool_list}, and its radius is proportional to the amount of speedup achieved by a drop-in replacement. See \figurename~\ref{f:clustering} for pattern definitions. To fully explain the different levels of speedup of different patterns, we need to consider not only randomness but also the other storage features. See the text for details. Discussions on ABySS, Maq, Bowtie 2, and BAW can be found in Section~\ref{s:discussion}. (\ba) Read. (\bb) Write. [\rsize, read size per request; \wsize, write size per request; \car, consecutive access ratio]}\label{f:randomness}
\end{figure*}

In \figurename~\ref{f:randomness}, for each of the two plots in this figure, the x-axis represents \car, while the y-axis corresponds to \rsize~or \wsize. For each of these features, recall from Section~\ref{ss:measuring_storage_features} that approaching 1.0 means that the access becomes more sequential, whereas going closer to 0.0 indicates more randomness in IO. Each program is represented by a circle, whose size is proportional to the amount of speedup by using SSDs. 

For the read case depicted in \figurename~\ref{f:randomness}(a), we see that the IO randomness, measured by either \rsize~or \car, is a reasonable first-order indicator for speedup. That is, either small \rsize~or \car~gives speedup by SSDs. For instance, the two patterns associated with steep speedup~(\pone~and \ptwo) manifest themselves through different types of randomness: \pone~has tiny \rsize~but its \car~is not small, whereas \ptwo~has small \car~but its \rsize~is high. \pfour~shows a typical sequential read behavior (both \rsize~and \car~are high), and the speedup is limited. Comparing \pfour~with \pone~or \ptwo~confirms that the read randomness is an important factor.

When both \rsize~and \car~have intermediate values, however, it is less obvious to predict the amount of speedup only by randomness. For instance, if we compare \pthree~and \pfive~in \figurename~\ref{f:randomness}(a) only by \rsize~and \car, then \pfive~should give higher speedup, which is not the case in reality. This is because the amount of IO is small for \pfive, as indicated in \figurename~\ref{f:clustering}(b) and (c), and there is little chance for SSDs to accelerate the IO.




In the write case depicted in \figurename~\ref{f:randomness}(b), we also observe that other storage features in addition to randomness need to be considered, although randomness remains an important factor for speedup. \ptwo~has small \car~and shows large speedup, which confirms that SSDs are effective for handling random writes. For the other patterns, we need to consider the role of write buffers inside storage devices. For writes, even HDDs can hide write latency to some extent using the write buffers. This can explain why \pfour~does not show speedup even though it has similar levels of randomness measured in \car~compared to \pone~or \pthree, both of which show noticeable speedup. \pone~and \pthree~ have higher \wsize~than \pfour, which leads them to have higher \bqlen.



\subsection{Impact of input size on SSD effectiveness}

We hypothesized that even tools that generate a small amount of IOs may benefit from using SSDs as the input size grows. Feeding large data may cause the main memory to be full generating frequent swaps. In this case, using SSDs may help reduce the runtime. 

\begin{figure}
\centering
\includegraphics[width=0.9\linewidth]{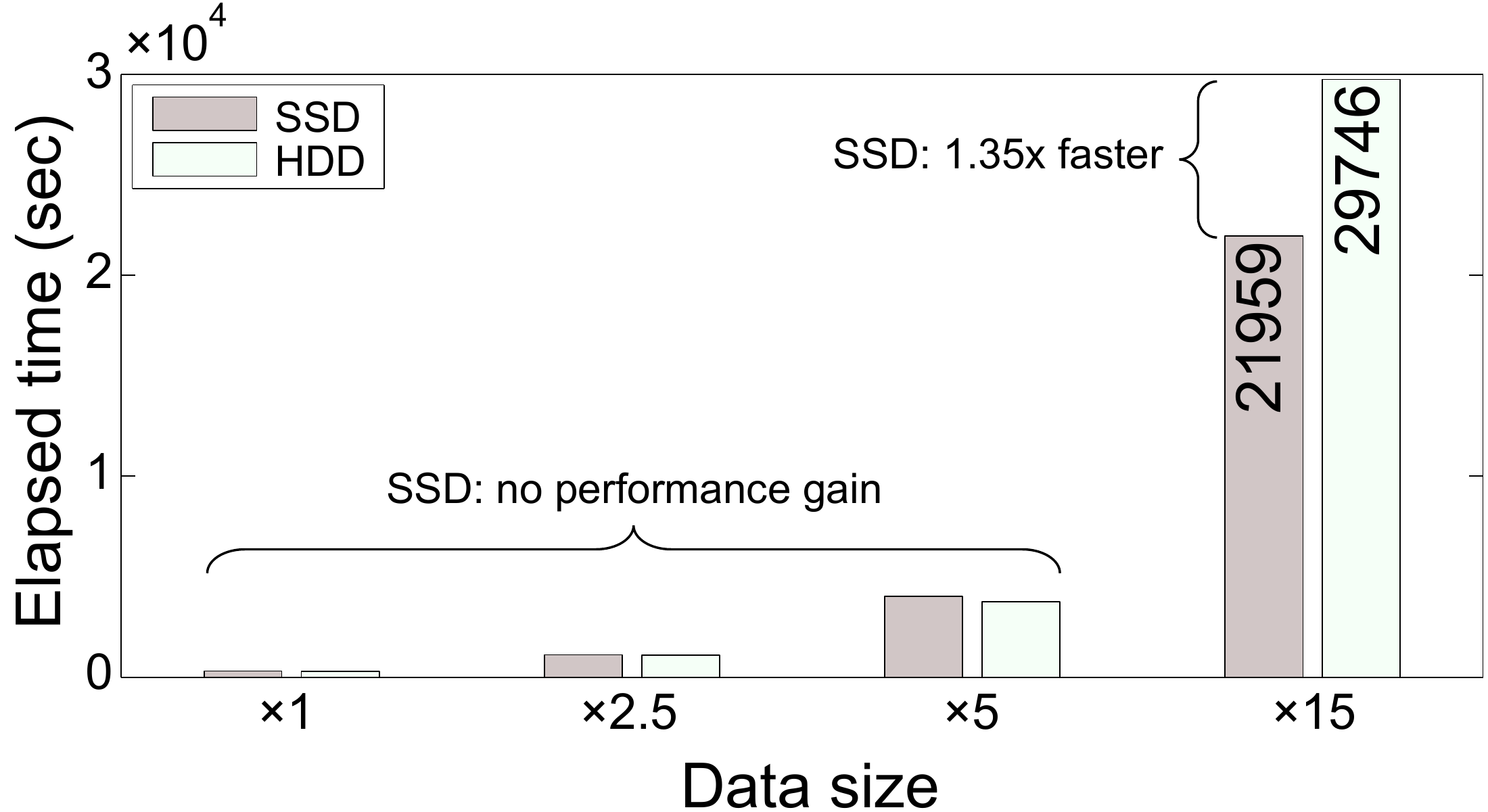}
\caption{CPU-intensive programs in \pfive~that produce a small amount of IO for moderate-size data may also benefit significantly by using SSDs for handling very large-scale data. [program: AmpliconNoise~\citep{quince2011removing}, baseline data: 2000 reads from 454 Titanium~\citep{quince2011removing}]}\label{f:amplicon_big}
\end{figure}

To verify this theory, we tried feeding increasingly larger data to AmpliconNoise~\citep{quince2011removing}, a program in \pfive. Recall that the programs in \pfive~are not very effectively accelerated by using SSDs, mainly because of their CPU-intensive behavior producing only a small amounts of IOs. The baseline data contains 2000 sequences sampled from the 454 Titanium data~\citep{quince2011removing}, and we generated larger data sets by replicating the baseline data. For each data set, we measured the runtime, as shown in \figurename~\ref{f:amplicon_big}. 

The breakeven point appears after replicating the baseline data five times. After that, using SSDs yields a huge speedup. This experiment confirms our theory and suggests that adopting SSDs may or may not be a smart decision, depending on the size of input data, even for the same program. For instance, AmpliconNoise often handles a number of pyrosequenced reads and is likely to benefit from using SSDs, although AmpliconNoise belongs to \pfive.


\subsection{Effect of main memory size on SSD-based acceleration}
The size of main memory affects the runtime of a workload, and ideally, the effect of using SSDs would be eclipsed in a system equipped with the main memory large enough for storing all the input/intermediate/output data. In reality, however, the memory footprint of a bioinformatics workload often becomes significantly larger than the main memory size affordable in typical systems, necessitating the use of a speedy secondary storage, such as SSDs.

We tested how the size of main memory affects the amount of speedup by SSDs using the GATK program, as shown in Figure 9. For an input dataset of 20GB sampled from the NA12878 human whole genome sequence [32], we ran the GATK using three sizes of main memory (4GB, 16GB, and 32GB) and measured the runtime of each of the four subprograms in the GATK for each memory configuration.

Using SSDs was most effective for the sequence base-quality recalibration (GATK BaseRecal) step, which shows high randomness in IO and belongs to P1. For the two memory sizes smaller than the input size (4GB and 16GB main memory), SSDs delivered a significant amount of speedup (66.32 times and 49.79 times, respectively). Even for the 32GB configuration, we observed more than 30 times speedup, which suggests that the memory footprint of GATK BaseRecal grows substantially during execution and the use of SSDs was effective.

For the sequence realignment step (GATK Aligner), the use of SSDs was helpful only for the 4GB memory configuration. For the setups with 16GB and 32GB memory, the amount of speedups was negligible. Although the input file size was 20GB, using SSDs was ineffective for 16GB main memory, which reveals the computing-intensive characteristic of the sequence alignment operation in GATK Aligner and the limited effectiveness of SSDs. For the other two programs (GATK Unified and GATK Printreads), we observed only negligible effects of using SSDs.

\subsection{Additional experimental results}\label{ss:additional}

In addition to the eight features listed in Table~\ref{t:io_features}, which are mostly related to storage devices, we measured CPU- and memory-related features (\eg, CPU usage and cache hit/miss ratios), as shown in \figurename~\ref{f:cpu_charac}. The CPU usage was higher for \gno, and the tools therein can be considered more compute-intensive than those in \gyes. The miss ratios for the lower-level caches and the translation lookaside buffer (TLB) tend to be higher for \gyes, confirming their memory-intensive behavior. The page fault rate was also higher for \gyes, which is compatible with the experimental results presented earlier.

\begin{figure}
\centering
\includegraphics[width=0.9\linewidth]{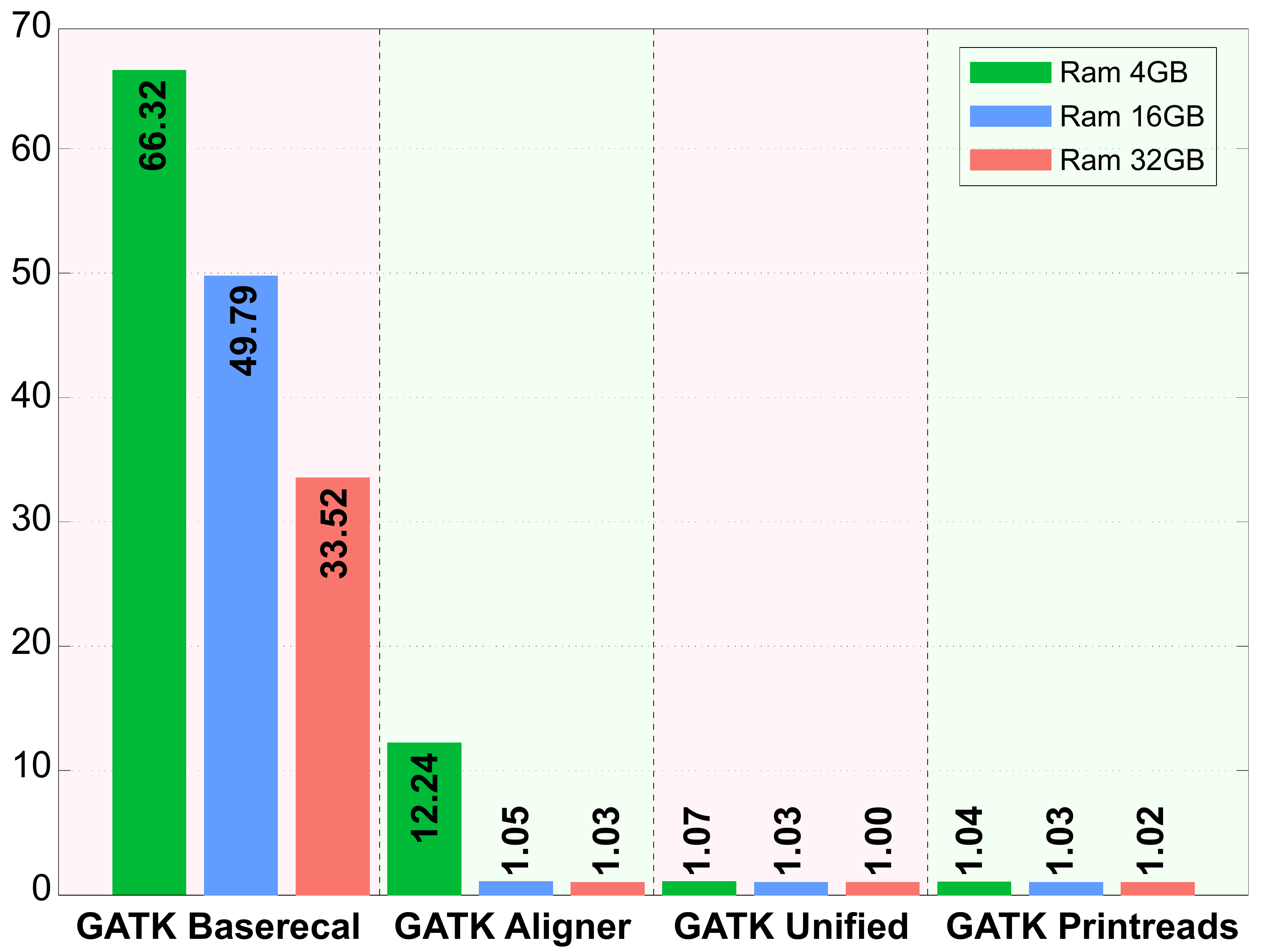}
\caption{Effects of main memory size on SSD-based acceleration of the GATK~\citep{mckenna2010genome}. Speedups of each of the four stages of the GATK with three different amounts of main memory are shown. [data: 20GB sample of NA12878 human whole genome sequence~\citep{10002010map}]
}
\end{figure}

\begin{figure}
\centering
\includegraphics[width=0.9\linewidth]{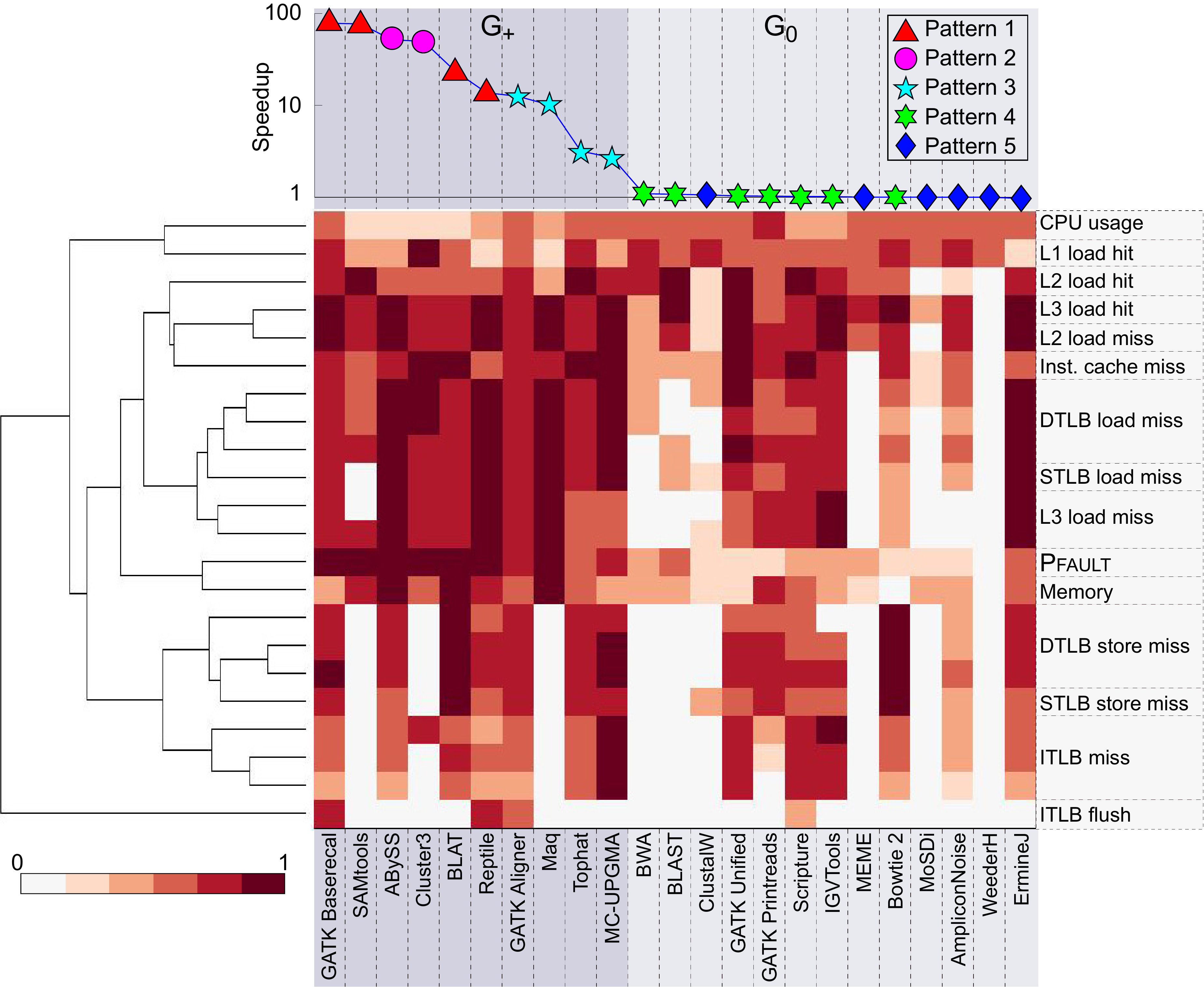}
\caption{Additional measurements of CPU- and memory-related features for the 23 bioinformatics programs. Speedup of each program is also shown. Features were normalized to values between 0 and 1. [\gyes, programs with 2x or more speedup; \gno, programs with negligible improvements]}\label{f:cpu_charac}
\end{figure}

\subsection{Summary and guidelines for employing SSDs in bioinformatics pipelines}

As seen in Figure 5(b) and 5(c), the most notable difference between \gyes~and \gno~ comes from the amount of page faults. In other words, when the memory footprint of a program exceeds the capacity of main memory, using SSDs is likely to bring a significant gain over using HDDs. By contrast, the programs with small memory footprints is less likely to be accelerated by using SSDs. 

Optimizing a program by reducing its memory footprint may bring a similar effect as using SSDs, but such a code optimization would typically require a nontrivial amount of efforts. Adopting SSDs thus becomes a more appealing option especially when the resources for code optimization are limited. Installing more main memory would also be helpful for reducing the runtime of programs, but the cost of DRAM may easily become prohibitively expensive, let alone the limited memory bandwidth issue.

Other factors that differentiate \gyes~and \gno~include the randomness of IO requests and the amount of data transfers: the more random and larger read/write requests, the more effective the use of SSDs. As the size of input data grows, even some of the programs in \gno~may benefit from using SSDs.

When deploying SSDs in a cluster environment, the administrator of the cluster should consider the network constraints before replacing HDDs with SSDs, because the effect of successful local acceleration may become eclipsed by the network latency, resulting in no overall performance gain (see Section 4 for more details).

\section{Discussion}\label{s:discussion}



The 23 programs we profiled represent traditional and emerging areas of importance, such as sequence alignment (including conventional dynamic programming-based, heuristic, and BWT-based algorithms), NGS denoising, assembly and mapping (including RNA-seq tools), gene expression analysis, motif finding, variant calling (including four GATK components), and metagenome analysis. These programs should cover the most frequent usages of bioinformatics data processing and related computation.

Through our experiments, we confirmed that acceleration by parallelization can be combined with the use of SSDs for even more performance increases. For example, using SSDs could accelerate ABySS more than 50 times, even though ABySS is a state-of-the-art parallelized assembler. The compute-intensive nature was mitigated by multicore processing, while the data-intensitve nature seems to have been handled by SSDs. The GATK package is another example. GATK was implemented using the map-reduce framework, which is amenable to parallel processing. In our experiments, SSDs could reduce the time demand of the two time-consuming components of GATK (BaseRecal and Aligner) by 78.4 and 12.6 times, respectively. When we design load balancing for parallelization, it will be helpful to consider the amount and randomness of IOs so that we can take advantage of SSDs.

In case the analysis pipeline contains a component program that is not accelerated by using SSDs, replacing the program with an alternative that runs faster on SSDs can help reduce the runtime of the overall pipeline. For example, in the sequence assembly and annotation pipeline depicted in \figurename~\ref{f:pipeline}(b), replacing Blast (only 1.08x speedup) with Blat (23.6x speedup) gave additional speedup to the whole pipeline. When there are multiple options for selecting a component block in a pipeline, it will thus be beneficial to assess the alternatives in terms of the effectiveness of using SSDs.


To this end, we can consider the three short-read aligners as an example: Maq (hash-based first-generation tool), Bowtie 2 and BWA (BWT-based second-generation tools). These three tools show similar \car~values, although Maq belongs to \pthree~and Bowtie 2 and BWA both belong to \pfour. In contrast, there is a difference in the IO size: Maq issues smaller reads but generates larger writes, which are linked to larger values of \pfault~and \bqlen. When HDDs are used, a critical limitation on the performance of Maq is put. To overcome this issue, significant efforts were made to invent the new generation of tools (Bowtie 2 and BWA) that have smaller memory footprints. The efforts could have been accompanied by using SSDs for even more improvements, given that the page faults and random IOs can be efficiently handled by SSDs.


There remain other intriguing topics for further research. A hybrid drive contains a spacious (but slow) HDD and a speedy (but small) SSD altogether inside a package. The access patterns are monitored, and frequently accessed ``hot'' data are cached automatically and dynamically in the SSD while the majority of the data are stored in the HDD. Using such a hybrid drive will be helpful for acceleration, under the conditions that the workload program creates enough IO requests (\eg, the programs in group \gyes) and the composition of the hot data do not change frequently over time. 

Exploiting the redundant array of independent disks (RAID) technology~\citep{hennessy2011computer} may provide additional advantages in performance and reliability. In particular, RAID level 0, which consists of striping without mirroring or parity, will be helpful for significantly improving data throughput. As long as the bandwidth of the host interface (\eg, SATA, PCIe, and NVM Express) is high enough to maintain the enhanced data throughput, using SSDs in RAID 0 will be helpful for accelerating high-throughput bioinformatics workloads.

Recently Hadoop-based clusters~\citep{taylor2010overview} are popular in large-scale data analytics including bioinformatics. The Hadoop file system (HDFS) provides a distributed storage layer on which various MapReduce-based operations are performed~\citep{chen2008map}. The randomness inherently occurring in the Map phase can be effectively handled by using SSDs~\citep{moon2014introducing}, which are far more superior to HDDs in terms of handling random IO requests. Improving the performance of a namenode (the node managing distributed file systems) in a Hadoop system by SSDs may provide another opportunity for SSD-based acceleration. In distributed systems, however, the network latency often eclipses the speedups achieved locally (\eg, shared-memory-based parallelization and SSD-based acceleration)~\citep{appuswamy2013scale}, and improving the overall performance globally may require significant efforts. Thus, even if the most frequently used applications in a cluster include the programs in the G+ group, the administrator of the cluster should carefully examine any network constraints that may exist before replacing HDDs with SSDs.


%
%
%
%
%
%
%
%


%

\section{Methods}\label{s:method}
%


\subsection{Experiment setup and measurements}\label{ss:exp_setup}

The SSDs and HDDs used in our experiments are listed in Table~\ref{t:ssd_spec} and Table~\ref{t:hdd_spec}, respectively. We selected these devices because they were the most popular in the market at the time of our experiments. For conservative comparison, the SSDs used are low-end models with 128GB or less capacity, whereas the HDD selection includes high-performance WD VelociRaptor.

Many of the bioinformatics tools we used take a long time to process large data especially when HDDs are used (often in the order of days or even weeks). To compare the performance of HDDs and SSDs using the same data sets while keeping experiments manageable, we selected, for each program, an input data set of appropriate size that can be processed in a reasonable amount of time (the criterion used: less than 72 hours). Table~\ref{t:data_list} lists details of the data used to profile the 23 bioinformatics programs.

To see the effects of changing secondary storages clearly in this setup, we also adjusted the specifications of the computer used accordingly. We used a machine equipped with a 3.3GHz Intel Core i3-3220 CPU (4 threads, 4MB L3 cache), 1600MHz dual-channel DDR3 memory (4GB for the GATK tools and 1GB for the others), and Ubuntu 12.04 LTS (Precise Pangolin).

For performance profiling and measurement, we used \texttt{time} (with option \texttt{-eUSKFW}), System Activity Reporter (SAR, \citealp{lu2013comprehensive}), blktrace~\citep{brunelle2007blktrace}, and Intel VTune Amplifier XE. To avoid interference between tools, we ran each of these profilers independently. We used \texttt{time} and SAR for measuring CPU usage and virtual-memory related features, blktrace for measuring block-level storage features (\eg, read/write amounts, throughput, and IOPS), and VTune for measuring CPU-internal features (\eg, cache hit/miss, TLB hit/miss, and IPC). When the range of measurements was large, we took the logarithm. We then normalized each of the measurements so that values were ranged in $[0,1]$. We repeated all the time measurements three times and used the average value for the analysis.



\begin{spacing}{\mytablespacing}
\ctable[
	caption = {Specifications of the SSDs used in this work},
	label = {t:ssd_spec},
]{lrrrrrr}{
}{
\toprule
\multirow{2}{*}{SSD}  & Capacity & \multicolumn{2}{c}{Sequential (MB/s)} & \multicolumn{2}{c}{Random (IOPS)}\\
  & (GB) & Read & Write & Read & Write\\
\midrule
Samsung 830 & 128 & 520 & 320 &80,000&30,000\\
Samsung 840 Pro & 128 & 530 & 390&97,000&90,000\\
OCZ Vertex4 & 128 & 560 & 430&90,000&85,000\\
Intel 520 & 120 & 550 & 475&50,000&80,000\\
Plextor M5 Pro & 128 & 540 & 330&91,000&82,000\\
Corsair Neutron GTX & 120 & 555 & 330 & 85,000&84,000\\
\bottomrule
}
\end{spacing}


\begin{spacing}{\mytablespacing}
	\ctable[
	caption = {Specifications of the HDDs used in this work},
	label = {t:hdd_spec},
]{lrrrrrrrr}{
}{
\toprule
\multirow{2}{*}{HDD}  & Capacity & \multirow{2}{*}{RPM} & Buffer size & Read/write &\multicolumn{2}{c}{IOPS (estimated)} \\
  & (GB) &  & (MB) & (MB/s) & Read&Write\\
\midrule
Seagate Barracuda & 1,000 & 7,200 & 64 & 156 &79.0 &73.2\\
WD Caviar Blue & 1,000 & 7,200 & 64 & 150&76.6 &66.4\\
WD VelociRaptor & 500 & 10,000 & 64 & 200&151.5 &138.9\\
\bottomrule
}
\end{spacing}






\begin{spacing}{\mytablespacing}
\ctable[
	caption = {List of the data used to test the 23 programs listed in Table~\ref{t:tool_list}},
	label = {t:data_list},
]{lll}{
\tnote[]{$\dagger$\tiny{\url{ftp://ftp-trace.ncbi.nih.gov/1000genomes/ftp/technical/working/20101201_cg_NA12878/}}; $\ddagger$\tiny{\url{http://trace.ddbj.nig.ac.jp/DRASearch/submission?acc=SRA012173}}; $\sharp$\tiny{\url{ftp://ftp.ncbi.nih.gov/pub/geo/DATA/supplementary/series/GSE20851/GSE20851_GSM521650_ES.aligned.sam.gz}};\hspace{0.5em}$\S$\tiny{\url{http://www.ncbi.nlm.nih.gov/geo/query/acc.cgi?acc=GPL92}}}
}{
\toprule 
Program  &Data & Source\\ 
\midrule 
GATK BaseRecal & NA12878 human & link$^\dagger$\\
Samtools & C2 & \citealp{bartram2011generation}\\
ABySS & \emph{Staphyloccus aureus} & \citealp{salzberg2012gage}\\
Cluster3 & Protein structure & \citealp{yoon2007clustering}\\
Blat & NCBI Uniref50 protein & \citealp{uniprot2010universal}\\
Reptile & Human chromosome 14 &\citealp{salzberg2012gage}\\
GATK Aligner&NA12878 human & link$^\dagger$\\
Maq &Human chromosome 14 &\citealp{salzberg2012gage}\\
Tophat &\emph{Drosophila melanogaster} & link$^\ddagger$ \\
MC-UPGMA & Protein structure & \citealp{yoon2007clustering}\\
\midrule 
BWA & AT1 & \citealp{bartram2011generation}\\
Blast & NCBI Uniref50 protein & \citealp{uniprot2010universal}\\
ClustalW& NCBI Uniref50 protein & \citealp{uniprot2010universal}\\
GATK Unified&NA12878 human & link$^\dagger$\\
GATK PrintReads&NA12878 human & link$^\dagger$\\
Scripture & Mouse (mm9) reads & link$^\sharp$\\
IGVtools& Mouse (mm9) reads & link$^\sharp$\\
Meme& Human sequence hm01 & \citealp{tompa2005assessing}\\
Bowtie 2&AT1 & \citealp{bartram2011generation}\\
Mosdi& Human sequence hm01 & \citealp{tompa2005assessing}\\
AmpliconNoise& 454 Titanium & \citealp{quince2011removing}\\
Weeder&Human sequence hm01 & \citealp{tompa2005assessing}\\
ErmineJ&Human genome U95 set & link$^\S$\\
\bottomrule
}
\end{spacing}

\subsection{More details of the storage features used}\label{ss:feature_exp}
Recall that we profile and analyze the 23 programs in terms of eight storage features that can characterize the amount and/or randomness of IOs.

To measure the amount of IOs we use three measures.
\butil~measures how much bandwidths of the interface between the host computer and the storage device are used. If there is a large amount of data transfers between the host and storage, \butil would be high.
\riops~and \wiops~measure how many read and write requests are made per second, respectively. A high value of these features implies frequent read/write requests.
\pfault~represents the number of page swaps per second. High \pfault~suggests frequent page swaps, which can be costly for HDDs.

The randomness of IOs can be measured in different ways. In this paper, we use two widely used measures: read/write size per request~\citep{li2014assert} and Consecutive Access Ratio (CAR,~\citealp{traeger2008nine}). Reads or writes that transfer a small amount of data are often considered random, whereas large read/write transfers are considered sequential. \car~measures how often consecutive accesses to the LBA space occur. The \car~value of one (zero) means perfectly sequential (random) IO access patterns.

\bqlen~represents the number of write requests waiting in the write buffer of a storage device. High \bqlen~normally can be caused by a high amount of write IOs and/or by a large number of small random writes. \bqlen~is thus related to both the amount and the randomness of IOs.

\section{Conclusion}\label{s:conclusion}
There exist cases in which a simple drop-in replacement of HDDs by SSDs can dramatically expedite bioinformatics programs. For instance, we observed more than 50 times speedup of widely used tools, such as GATK components, Samtools, and ABySS. In the arena of short-read aligners, we observed that Maq (a hash-based first-generation tool) could compete again with Bowtie 2 and BWA (the second-generation tools) leveraged by SSDs. According to our experiments, using SSDs could accelerate the GATK-based variant calling pipeline by more than 30 times.

However, SSDs are not silver bullets and cannot boost every bioinformatics program of one’s interest. Moreover, SSDs are still expensive. Eventually the price of SSDs may become competitive to HDDs, but the price per gigabyte of SSDs is still approximately 15 times more expensive, as of 2015. Researchers handling large-scale biomedical data should thus make a careful and informed decision regarding whether to replace their HDDs (at least partially) with SSDs.

To this end, profiling the bioinformatics tools of interest from system perspectives is critical. According to our experiments, there exist many bioinformatics programs that can benefit immediately by using SSDs, especially when the program causes frequent random IOs or page swaps due to relatively large input compared to system memory. This review reports other patterns indicating the viability of SSD-based acceleration. As the size of input data grows, we expect that the territory of the SSD-acceleratable programs will expand.

In any case, as the performance of SSDs is rapidly improving with continuous cost reduction and technology developments, SSDs will eventually become the storage device of choice, phasing out HDDs firstly in performance-critical domains and later in the mainstream. We thus believe that future bioinformatics algorithms should be designed to consider the advantage of using SSDs in addition to the applicability of parallel processing. We hope that the results and insight presented in this review will be a valuable asset to such a journey for inventing efficient and scalable bioinformatics tools.

\section*{Acknowledgements}
The authors would like to thank Byunghan Lee, Sunyoung Kwon and Sei Joon Kim at Yoon Lab for helpful discussion.

This work was supported by the National Research Foundation (NRF) of Korea grants funded by the Korean Government (Ministry of Science, ICT and Future Planning) [No. 2011-0009963, No. 2014M3C9A3063541]; the ICT R\&D program of MSIP/ITP [14-824-09-014, Basic Software Research in Human-level Lifelong Machine Learning (Machine Learning Center)]; SNU ECE Brain Korea 21+ project in 2015; and Samsung Electronics Co., Ltd.


\begin{spacing}{1}
\bibliographystyle{ieeetr}
\bibliography{reference,test}
\end{spacing}

\end{spacing}
\end{document}